\documentclass[conference]{IEEEtran}

\usepackage{moreverb}
\usepackage[T1]{fontenc}
\usepackage{hhline}
\usepackage{amssymb,amsfonts,amsthm}
\usepackage{amsmath}
\usepackage{amsfonts}
\usepackage{epsfig}
\usepackage{color}
\usepackage{enumerate}
\usepackage{ifpdf}
\usepackage{graphicx}
\usepackage{cite}
\usepackage{subfig}
\usepackage{caption}
\usepackage{multirow}
\usepackage[table,xcdraw]{xcolor}
\usepackage{mathtools}

\newtheorem{mytheorem}{Theorem}
\newtheorem{mycor}{Corollary}
\newtheorem{myprop}{Proposition}

\begin{document}

\title{Geo-Network Coding Function Virtualization for Reliable Communication
over Satellite}

\author{Tan Do-Duy, M. Angeles Vazquez Castro\\
Dept. of Telecommunications and Systems Engineering\\
Autonomous University of Barcelona, Spain\\
Email: \{tan.doduy, angeles.vazquez\}@uab.es}

\maketitle
\begin{abstract}
In this paper, we propose a design solution for the implementation
of Virtualized Network Coding Functionality (VNCF) over a service
coverage area. Network Function Virtualization (NFV) and Network Coding
(NC) architectural designs are integrated as a toolbox of NC design
domains so that NC can be implemented over different underlying physical
networks including satellite or hybrid networks. 

The design includes identifying theoretical limits of NC over wireless
networks in terms of achievable rate region and optimizing coding
rates for nodes that implement VNCF. The overall design target is
to achieve a given multicast transmission target reliability at receiver
sides. In addition, the optimization problem uses databases with geo-tagged
link statistics and geo-location information of network nodes in the
deployment area for some computational complexity/energy constraints. 

Numerical results provide validation of our design solution on how
network conditions and system constraints impact the design and implementation
of NC and how VNCF allows reliable communication over wireless networks
with reliability and connectivity up to theoretical limits.
\end{abstract}

\section{Introduction \label{sec:intro}}

NC has attracted much attention in recent years as key concepts for
5G networks to provide flexibility and substantial gains in throughput
and reliability. The performance of NC depends on the deployment of
special nodes on the path, known as the coding points, that perform
re-encoding the coded packets from the source in order to allow capacity
achievability both for noiseless \cite{Ahlswede.2000} and noisy networks
\cite{Dana.2006}.

As an innovative technique towards the implementation of network functions,
NFV is clearly a NC design option as it would make NC available as
a flow engineering functionality offered to the network. NFV has been
proposed as a promising design paradigm by the telecommunications
sector to facilitate the design, deployment, and management of networking
services. Essentially, NFV separates software from physical hardware
so that a network service can be implemented as a set of virtualized
network functions through virtualization techniques and run on commodity
standard hardware.

The integration of NC and NFV will enable the applicability of NC
in future networks (e.g. upcoming 5G networks) to both distributed
(i.e. each network device) and centralized manners (i.e. servers or
service providers). The European Telecommunications Standards Institute
(ETSI) has proposed some use cases for NFV in \cite{NFV001.2013},
including the virtualization of cellular base station, fixed access
network, etc. There are already available proposals that combine NC
virtualization and software-defined networking (SDN). For example,
in \cite{Szabo2.2015}, authors investigate the usability of random
linear NC as a function in SDN to be deployed with virtual (software)
OpenFlow switches. In \cite{Hansen.2015}, NC is implemented in a
virtual machine which is then embedded into an Open vSwitch. These
works indicate the feasibility of integrating NC as a functionality
based on SDN and the concentration of network functions in centralized
architectures such as data centers or centralized locations proposed
by network operators and service providers. In addition, it has been
shown in \cite{Bertaux.2015} benefits of SDN and virtualization to
satellite networks as well as their impacts on a typical satellite
system architecture. However, a unified design framework for NC design
in view of NFV either centralized or distributed is currently missing.

Our motivation is the generalization of NC design domains so NC can
be applied over different operational services including satellite
or hybrid networks thus enabling flexible deployment, which is expected
to derive significant benefits for communication over satellite. Furthermore,
databases with geo-tagged link statistics and geo-location information
also provide useful information for the NC optimization functionality.
Our contributions can be summarized as follows:
\begin{itemize}
\item We propose an architectural design framework for NC and integration
of NC and NFV architectural design as a toolbox of NC domains.
\item As part of coding design domain, we analyze theoretical limits of
NC over wireless networks in terms of achievable rate region for some
target residual erasure rates.
\item Validation of the proposed VNCF design for a complete use case to
increase reliability and connectivity for communication service over
satellite network. Specifically, we clearly identify general procedure
for the instantiation, monitoring, and termination of VNCF. Also,
the design optimizes coding rates according to geo-tagged link statistics
and complexity/energy constraints.
\end{itemize}
The rest of this paper is organized as follows. In Section \ref{sec:DESIGN},
we propose the architectural design of NC. Section \ref{sec:CodingDomain}
presents coding design domain including NC model and analysis of per-receiver
achievable rate region. In Section \ref{sec:USE_CASE}, we validate
our proposed architectural design in a complete design. Several numerical
results are conducted in Section \ref{sec:SIM} to validate the performance
for the case of using our NC design. Finally, Section \ref{sec:CONC}
identifies conclusions and further work.

\section{NC Architectural design domains\label{sec:DESIGN} }

The combination of NFV and NC brings forth a potential solution for
the management and operation of the future networks. NC design involves
different domains \cite{MAVazquez2.2015}: 
\begin{itemize}
\item \textbf{NC coding domain} - domain for the design of network codebooks,
encoding/decoding algorithms, performance benchmarks, appropriate
mathematical-to-logic maps, etc.
\item \textbf{NC functional domain} - domain for the design of the functional
properties of NC to match design requirements built upon abstractions
of 

\begin{itemize}
\item \textbf{Network}: by choosing a reference layer in the standardized
protocol stacks and logical nodes for NC and re-encoding operations. 
\item \textbf{System}: by abstracting the underlying physical or functional
system at the selected layer e.g. SDN and/or function virtualization. 
\end{itemize}
\item \textbf{NC protocol domain} - domain for the design of physical signaling/transporting
of the information flow across the virtualized physical networks in
one way or interactive protocols. 
\end{itemize}
In next section, we introduce systematic NC (SNC) as part of coding
design domain and then analyze per-receiver achievable rate region.

\section{Coding Design Domain \label{sec:CodingDomain} }

\subsection{Network Coding model}

We consider a multi-hop line network. Let $\delta_{i}$ be erasure
rate of each link $i$ and $\overline{\delta}$ be the vector of per-link
erasure rates e.g. $\overline{\delta}$ = ($\delta_{1}$, $\delta_{2}$)
for 2 hops. If all links are equal, we use $\delta$. Assume that
a source sends $k$ data packets followed by $n-k$ random linear
combinations with coefficient matrix generated using coefficients
from the same Galois field of size $g=2^{q}$. Then, coding rate is
given by $r=k/n$ for some target residual erasure rate at the receivers,
$\eta_{0}$. We also assume that packet length is $L$ bits, resulting
in $s=L/q$ symbols per packet. 

Shrader et al. \cite{Shrader.2009} indicate that if SNC is used at
cut-node(s), then network throughput will not reduced relative to
random linear NC. Moreover, SNC has the benefits of lower computational
complexity needed to construct random linear combinations and decode
the received packets at the destinations. Therefore, SNC should be
a candidate for NC codebook. We denote $\eta_{i}\left(r,\delta_{i}\right)$
as the residual packet erasure rate (RPER) after decoding at each
single hop $i$. Analytical expressions of RPER with SNC are derived
in \cite{Saxena.2016}.

In view of multi-hop line networks, we define the reliability after
decoding at hop $h^{th}$ as
\begin{equation}
\rho_{R}^{NC}(r,\overline{\delta},h)=\text{\ensuremath{\prod}}_{i=1}^{h}\left(1-\eta_{i}\left(r,\delta_{i}\right)\right).
\end{equation}

The achievable rate of a receiver located at hop $m$ is defined as
$R^{m}=r(1-\eta^{m})$ with $\eta^{m}=1-\text{\ensuremath{\prod}}_{i=1}^{m}\left(1-\eta_{i}\left(r,\delta_{i}\right)\right)$
is RPER at hop $m$. 

Accordingly, we also define the reliability at hop $h^{th}$ for the
uncoded case as
\begin{equation}
\rho_{R}^{noNC}(\overline{\delta},h)=\text{\ensuremath{\prod}}_{i=1}^{h}\left(1-\delta_{i}\right).
\end{equation}

If compared with using spatial diversity in which $n$ parallel nodes
without re-encoding in between the source and the end node carrying
the same coded packets, using one line network with re-encoding at
intermediate node can provide a decoding probability reaching the
performance of the diversity case but with lower network resource
and complexity of management.

\subsection{Per-receiver Achievable Rate Region}
\begin{mytheorem}
For multi-hop line networks with $V$ vertices and an arbitrary receiver
$m$ $(m\in[2,|V|])$, the achievable rate region per receiver with
SNC is the set of points is located within the polytope in $R^{|V|}$
for different target RPER, \textup{$\eta_{0}$,} and determined by
the following inequalities: 
\end{mytheorem}
\begin{eqnarray} 
R^{m}	& \leq & min_{i:\,i\in[1,m-1]}(1-\delta_{i}),\\ 
\eta^{m} & \leq & \eta_{0},\\ 
R^{m}	& \leq & R^{m-1}.
\end{eqnarray}
\begin{IEEEproof}
See Appendix 1.
\end{IEEEproof}
\begin{mycor}
At an arbitrary point at the capacity, i.e. $\eta_{0}=0$ is required,
$R^{m}=R^{m-1}=r$ \textup{~$(\forall m\in[2,|V|])$}. While at an
arbitrary point in the achievable rate region and the target $\eta_{0}\neq0$,
$R^{m}<R^{m-1}$ \textup{~$(\forall m\in[2,|V|])$}.
\end{mycor}
Conclusive results inferred from Theorem 1 are as follows:
\begin{itemize}
\item We only need to optimize coding rate for the sink/final destination
which is located at the end of the line. If the target requirement
of the sink is satisfied, the target requirement of the intermediate
nodes is also satisfied.
\item The larger the target $\eta_{0}$, the better the decoding performance
that the intermediate nodes obtain if compared with that of the sink.
\end{itemize}

\section{Case Study: Geo-Network Coding Function Virtualization \label{sec:USE_CASE}}

We propose a complete Virtualized Geo-NC Function (VGNCF) design for
reliable communication services over satellite, called \textit{geo-controlled
network reliability}. In the illustrative design proposed here, the
optimization functionality will use databases with geo-tagged link
statistics and geo-location information of nodes in the deployment
area. The overall design target here is to achieve a given connectivity
and/or target reliability throughout the service coverage area for
some complexity constraints.

\begin{figure}[tbh]
\begin{centering}
\includegraphics[scale=0.50]{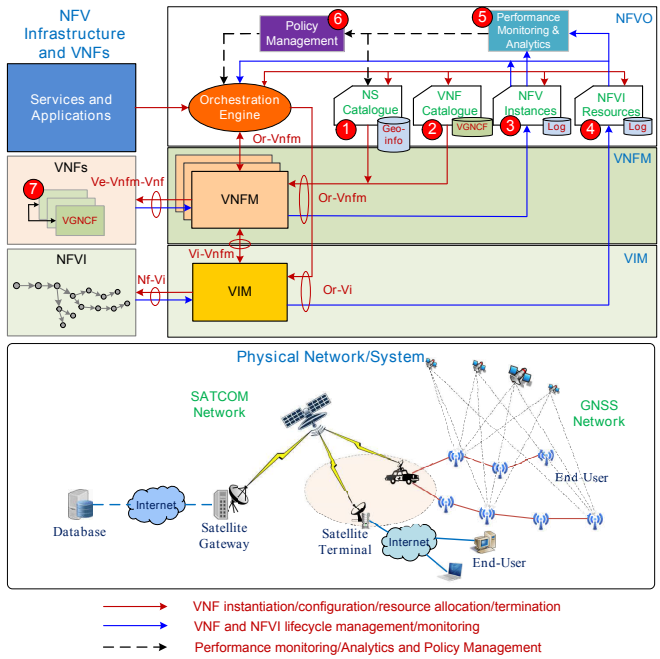}
\par\end{centering}
\caption{Integration of VGNCF and ETSI NFV architecture \cite{NFV002.2013}
and exchanges between VGNCF and NFV-MANO blocks via reference points.
\label{fig:EXCHANGES}}
\end{figure}

\subsection{Physical system/network abstraction \label{sec:Abstraction}}

The interest of NFV is that the same virtualized network function
(VNF) can be applied over different operational frameworks and services
as well as over different underlying physical networks, including
satellite or hybrid networks thus enabling rapid innovative deployment.
In Fig. \ref{fig:EXCHANGES}, we abstract the underlying network to
identify NC functionalities. All nodes in the deployment area are
geo-localized based on GNSS network. Our proposed use case applies
to multiple scenarios consisting of terminals (randomly) distributed
in the service area both within and beyond cell coverage of satellites.
Connectivity to/from the source is realized via satellites and/or
transport networks (e.g. internet, cellular networks), depending on
availability but in all cases they are abstracted for our design as
sender/sink (end-user) nodes, respectively, through a virtualization
layer.

\subsection{Virtualized Geo-NC function architecture \label{sec:GeoNC_Architecture}}

Fig. \ref{fig:EXCHANGES} shows how VGNCF can be integrated with the
ETSI NFV architecture given the abstracted underlying physical system/network.
The design also indicates exchanges between VGNCF and NFV-Management
and Orchestration (NFV-MANO) over reference points. Specifically,
NFV-MANO includes three functional blocks: NFV Orchestrator (NFVO),
VNF Manager (VNFM), and Virtualized Infrastructure Manager (VIM).
NFVO block has two main responsibilities: (1) orchestration of NFV
Infrastructure (NFVI) resources across multiple VIMs and (2) life-cycle
management of all network services. While VNFM manages the life-cycle
of VNF instances, VIM is responsible for managing and controlling
NFVI resources including physical, virtual, and software resources.
More details of the NFV-MANO architecture, reference points and repositories
can be found in \cite{NFV-MAN001.2014}. 

\begin{figure}[tbh]
\begin{centering}
\includegraphics[scale=0.55]{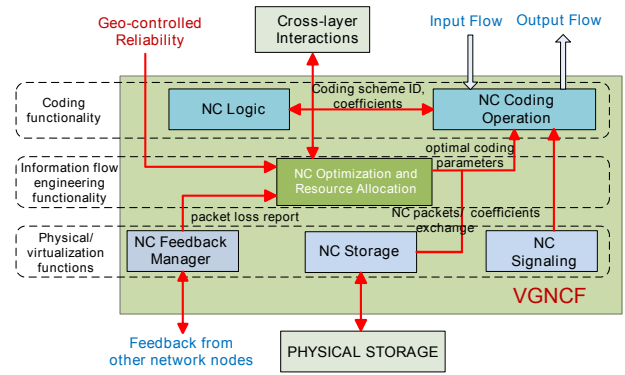}
\par\end{centering}
\caption{Functional domain design of VGNCF following our proposed NC architectural
design framework. \label{fig:NC_Functional}}
\end{figure}

As denoted in Fig. \ref{fig:EXCHANGES}, the management and orchestration
of VGNCF is highlighted by the following points:
\begin{enumerate}
\item NS catalogue holds information of all usable NSs in terms of VNFs
and description of their connectivity through virtual links. Geo-information
e.g. geo-tagged link statistics and geo-location information is also
stored in NS catalogue.
\item VNF catalogue holds information of all usable VNFs in terms of VNF
Descriptors.
\item NFV instances repository keeps record from VGNCF during its execution
of management operations and life-cycle management operations.
\item NFVI resources hold information about NFVI resources utilized for
VNF/NS operations.
\item Performance monitoring and analytics block takes care of collecting
and analyzing non-real-time/real-time data e.g. loss rate resulting
from operation of VGNCF.
\item Policy management receives information from performance monitoring
and analytics block to decide whether coding policy should be changed
or not.
\item Interaction between VNFs should be considered. For instance, a virtualized
routing function implemented provides route information to VGNCF.
\end{enumerate}
The detailed functional domain of our proposed VGNCF is shown in Fig.
\ref{fig:NC_Functional}. At the coding functionality blocks, interactions
with other nodes bring into agreement on coding schemes, coefficients
selection, etc. NC coding operation block receives all inputs such
as coding scheme, coefficients, coding parameters, packets from storage
block, and signaling to perform elementary encoding/re-encoding/decoding
operations, etc. At the information flow engineering functionality
blocks, geographical location-based information and level of reliability
provided by \textit{geo-controlled reliability} block will be given
to the NC optimization and resource allocation block so that optimal
coding parameters are generated to the NC coding operation block.
In addition, packet loss feedback from other network nodes is also
an important factor in the resource allocation process. At the last
stage, the physical virtualization functions blocks, which connect
directly to physical storage and feedback from other network nodes
to provide information packets and packet loss rate to the upper stage,
respectively.

\subsection{Procedure for the instantiation/monitoring/termination of VGNCF}

\begin{figure*}[tbh]
\begin{centering}
\includegraphics[scale=0.80]{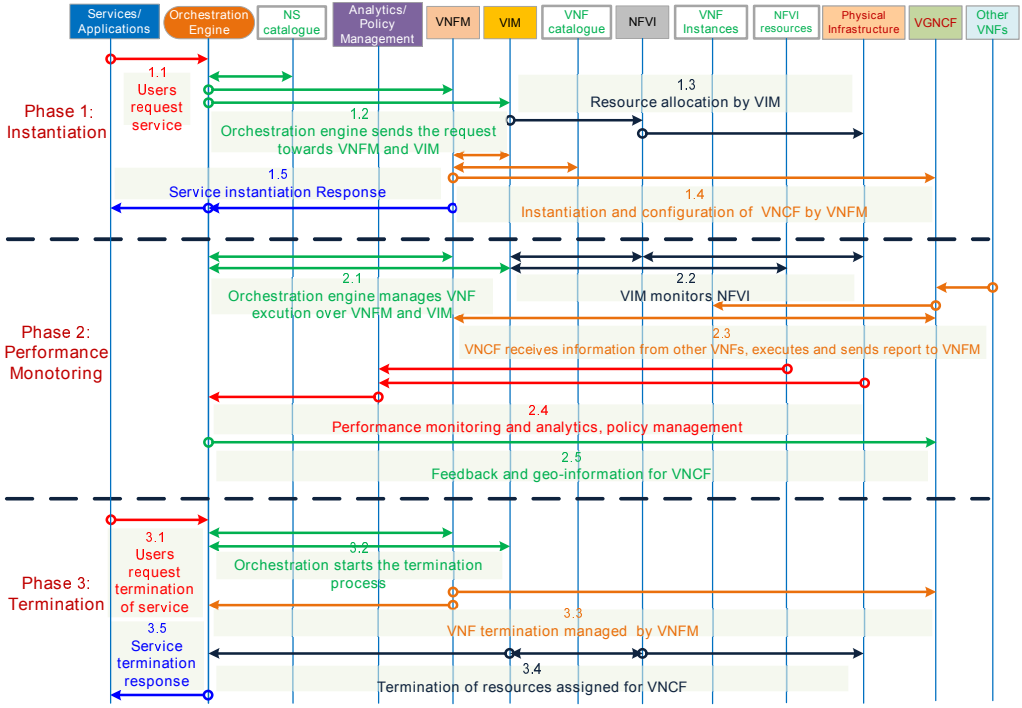}
\par\end{centering}
\caption{Procedure for the instantiation, execution and monitoring, and termination
of VGNCF. \label{fig:PROCEDURE}}
\end{figure*}

In Fig. \ref{fig:PROCEDURE}, we denote a general procedure of the
instantiation, performance monitoring, and termination phases corresponding
life-cycle management of VGNCF execution. The details are as follows:
\begin{itemize}
\item \textit{Phase 1.1:} customers/users request the reliability functionality
for a communication service over satellite.
\item \textit{Phase 1.2:} based on the descriptor received from the NS catalogue,
the Orchestration Engine (OE) delivers a message towards the VNFM
and VIM to request for the instantiation of VGNCF and resource allocation.
\item \textit{Phase 1.3: }VIM takes charge of managing and allocating resource
required for the instantiation and execution of VGNCF via NFVI. The
underlying physical infrastructure would be allocated respectively.
\item \textit{Phase 1.4:} VNFM interacts with VNF catalogue for the description
of VGNCF in terms of its deployment and operational behavior and delivers
necessary configuration for the instantiation of VGNCF at coding points.
\item \textit{Phase 1.5:} the instantiation phase has been done by acknowledgment
from VNFM to OE and the user in order to confirm the successful deployment
of VGNCF.
\item \textit{Phase 2.1:} during execution of VNF/NS life-cycle management
operation, the OE keeps interacting with VNFM and VIM for real-time
management and monitoring.
\item \textit{Phase 2.2:} VIM takes care of NFVI and network resources while
current information of NFVI resources utilized for VNF/NS operations
is stored in NFVI resources catalogue for future use and ready for
using by the OE.
\item \textit{Phase 2.3:} VGNCF receives routing information from another
VNF, e.g. virtualized routing function, and configuration from VNFM
for executing coding functionality.
\item \textit{Phase 2.4:} the performance monitoring/analytics and policy
management blocks gather information from real-time/non real-time
operation of VGNCF recorded in NFV instances and NFVI resources for
analytical framework. The output is then sent to the OE in order to
update configuration and coding policy.
\item \textit{Phase 2.5:} VGNCF requires updating feedback and geo-information
for optimization functionality.
\item \textit{Phase 3.1:} assume that users would like to deactivate NC
functionality. Then, a request is sent to the OE.
\item \textit{Phase 3.2:} OE delivers the termination message for the VNFM
and VIM.
\item \textit{Phase 3.3:} VNFM deactivates all functionalities of VGNCF.
\item \textit{Phase 3.4:} corresponding resources provided for VGNCF will
be deallocated and released back to the VIM.
\item \textit{Phase 3.5:} after receiving the termination response from
both VNFM and VIM, the OE acknowledges completion of the termination
phase to central controller/users.
\end{itemize}

\subsection{Optimization functionality\label{sec:OPTIMIZATION}}

\subsubsection{Computational Complexity\label{sec:Complexity}}

Let $\beta_{0}^{S}$, $\beta_{0}^{R_{j}}$, and $\beta_{0}^{D}$ denote
limitations on computational complexity of each node implementing
VGNCF for the source, the intermediate nodes, and the final destination,
respectively. Let $\beta_{S}(r)$, $\beta_{R_{j}}(r)$, and $\beta_{D}(r)$
denote computational complexity required for coding operation at the
source, re-encoding points $R_{j}$, and the destination, respectively.
For more precise evaluation, we consider the computational complexity
in terms of total number of logic gates required for implementing
multiplication and addition operations over $GF(2^{q})$ which can
be approximated by $q$-bit arithmetic operations as $2q^{2}+2q$
and $q$ logic gates, respectively \cite{Angelopoulos.2011}. 
\begin{myprop}
The number of multiplications and additions required for encoding
process is $N_{enc}^{M}=(n-k)ks$ and $N_{enc}^{A}=(n-k)(k-1)s$,
respectively. Therefore, computational complexity for encoding in
terms of logic gates is $\beta^{enc}(r)=N_{enc}^{M}(2q^{2}+2q)+N_{enc}^{A}q$.
Whereas computational complexity of finite-length decoding complexity
of Gaussian Elimination algorithm is denoted by $\beta^{dec}(r)=N_{dec}^{M}(2q^{2}+2q)+N_{dec}^{A}q$
with \textup{$N_{dec}^{M}$} and \textup{$N_{dec}^{A}$} are given
in \textup{\cite{Garrammone.2013}}.
\end{myprop}
Each re-encoding point $j$, as a receiver on the path, will decode
and re-encode the linear combinations before forwarding the coded
packets towards next hops. Without decoding the re-encoder does not
know which packets are innovative, while with decoding the relay may
do a more intelligent re-encoding operation. Therefore, the complexity
required for the relay is the total of complexity for decoding and
re-encoding, i.e. $\beta_{R_{j}}(r)=\beta^{dec}(r)+\beta^{enc}(r)$,
whereas $\beta_{S}(r)=\beta^{enc}(r)$ and $\beta_{D}(r)=\beta^{dec}(r)$.
Assume that coding rate is same for all nodes.

\subsubsection{Utility Function\label{sec:Utility}}

We are interested in the design of when NC should be activated in
terms of (1) computational complexity and (2) target reliability after
decoding $(\rho_{0})$. In order to do so, we define the following
utility function: 

\begin{equation}
u^{act}(r,\overline{\delta},\rho_{0})=\frac{f^{NC}(r,\overline{\delta},\rho_{0})}{f^{COST}(r)},
\end{equation}

where $f^{NC}(r,\overline{\delta},\rho_{0})$ accounts for the goodness
of the coding scheme in achieving target performance $\rho_{0}$.
Whereas $f^{COST}(r)$ accounts for the cost in terms of computational
complexity. The utility function denotes the efficiency in terms of
the ratio between the goodness and the computational complexity. We
define the goodness and the cost function, respectively as follows 

\begin{equation}
f^{NC}(r,\overline{\delta},\rho_{0})=\rho_{R}^{NC}(r,\overline{\delta})-\rho_{0},
\end{equation}

\begin{equation}
f^{COST}(r)=\beta_{S}\left(r\right).
\end{equation}

\subsubsection{Optimized operative range of performance\label{sec:Optimization}}

Underlying geo-tagged channel statistics are assumed to be stored
in NS catalogue (see Fig. \ref{fig:EXCHANGES}). The upper bound for
energy-efficient coding rate is given by the following proposition.
\begin{myprop}
The source identifies at which rate it maximizes its own utility under
constraints of computational complexity by the following strategy: 
\end{myprop}
\begin{equation} 
\begin{aligned} 
& \underset{r}{\text{argmax}} 
& & u^{act}(r,\overline{\delta},\rho_{0}) \\ 
& \text{subject to} 
& & \beta_{S}(r)\le\beta_{0}^{S},\\ 
& & & \beta_{D}(r)\le\beta_{0}^{D},\\ 
& & & \beta_{R_j}(r)\le\beta_{0}^{R_j}. 
\end{aligned} 
\label{eq:OPT} 
\end{equation} 

Numerical results reveal that the utility function has the property
of quasi-concavity. Moreover, $\beta_{D}(r)$ and $\beta_{R_{j}}(r)$
are increasing functions with redundant coded packets. Therefore,
Problem (9) is equivalently quasi-convex optimization and can be efficiently
solved by bisection methods. The utility may hold a minus value which
represents the penalty due to the dissatisfaction of the design target.
Assume computational complexity limitation is large enough, the optimal
utility is not necessarily at coding rates that make $\rho_{R}^{NC}(.)=1$. 

It is necessary to identify optimized operative ranges of performance
so that the destination is aware and admits some variations in the
quality. At the source or a centralized controller, the cognitive
algorithm to identify optimized operative ranges is briefly realized
as follows: 

(1) identify maximal utility $u_{max}^{act}(r,\overline{\delta},\rho_{0})$
and respective $r$, $\rho_{R}^{NC}(r,\overline{\delta})$ given $(\overline{\delta},\rho_{0})$, 

(2) determine $r$ and $\rho_{R}^{NC}(r,\overline{\delta})$ that
satisfies $u_{min}^{act}(.)\leq u^{act}(.)\leq u_{max}^{act}(.)$,
with $u_{min}^{act}(.)$ is the lower bound, and 

(3) activate NC functionality if the range of performance is acceptable
by users. 

Note that we can assume the lower bound is just the smallest value
of utility that the target $\rho_{0}$ is still satisfied. For the
sack of comparison, we also assume that even though the target cannot
be met for any coding rate due to dramatic erasure process and/or
low complexity constraint, the source would show its effort to improve
network reliability as much as possible regardless its utility may
still be minus. 

\section{Numerical Results \label{sec:SIM}}

\subsection{Per-link achievable rate region \label{sec:Perlink_RATE}}

In this section, we indicate that by adding redundancy not only at
the source but also at intermediate nodes, per-link achievable rate
region goes beyond that of the case that NC is applied only at the
source. We consider a single-source multicast network, e.g. satellite
network \cite{MAVazquez2.2015}. We denote the first case as \textit{NC-cas}e
where VGNCF is applied at both the source and the intermediate node.
While the second one is with NC at the source only and known in the
NC literature as \textit{end-to-end coding} to which NC can be compared
with since the erasure seen by the encoder is the total erasure in
the network. It is noted that we use interchangeably the target reliability
$\rho_{0}$ and $\eta_{0}$, with $\eta_{0}=1-\rho_{0}$. Let denote
$R^{NC}=r\left(1-\eta^{NC}\right)$ and $R^{e2e}=r\left(1-\eta^{e2e}\right)$
as the achievable rate for NC case and end-to-end coding, respectively,
where $\eta^{NC}$ is RPER for a two-hop network and $\eta^{e2e}$
corresponds RPER for a single-hop network but with erasure rate is
total erasure of two hops. 

\begin{figure}[tbh]
\subfloat[$R^{e2e}$ ]{\includegraphics[scale=0.36]{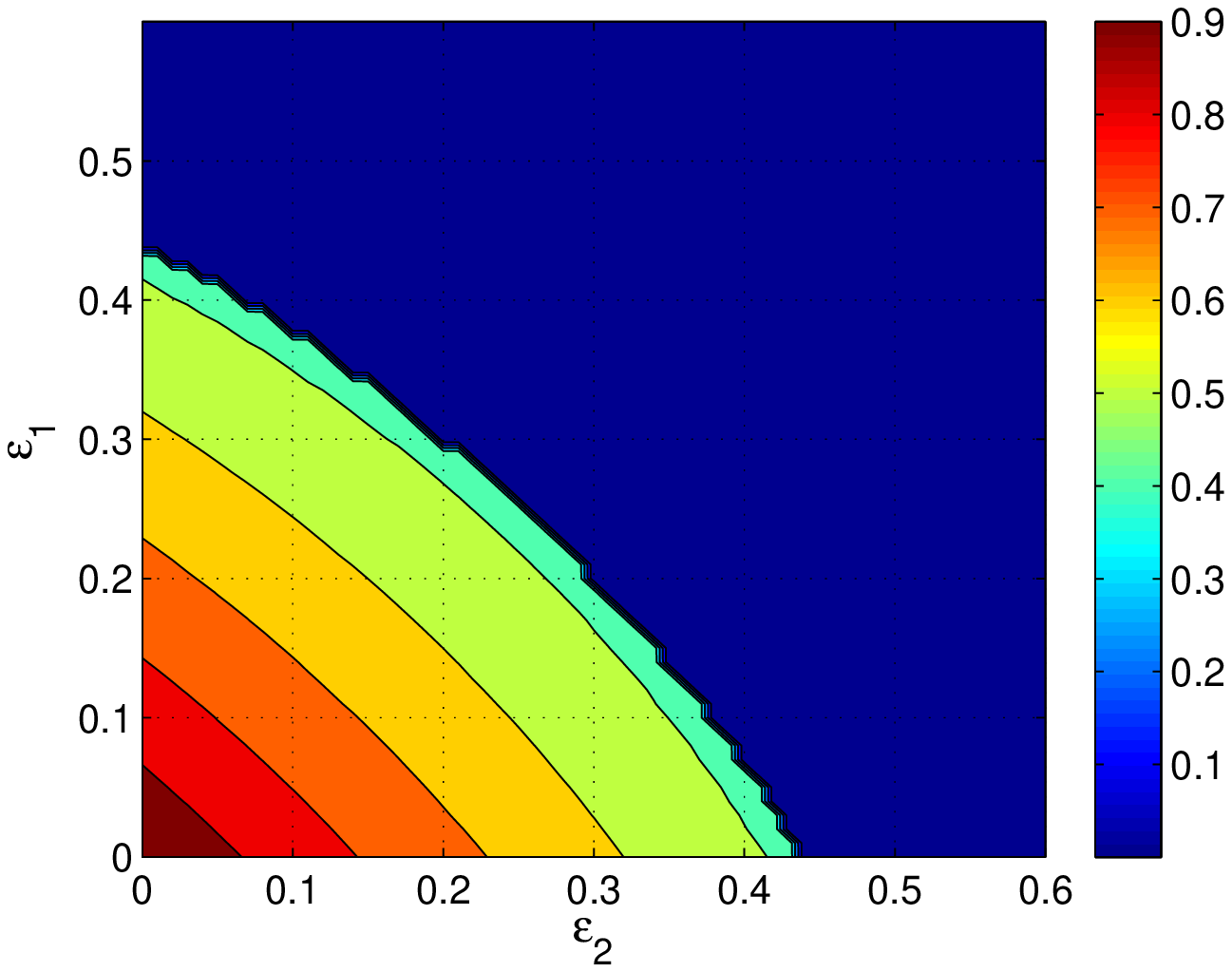}

\label{fig:R2e2}}\ \ \subfloat[$R^{NC}$]{\includegraphics[scale=0.36]{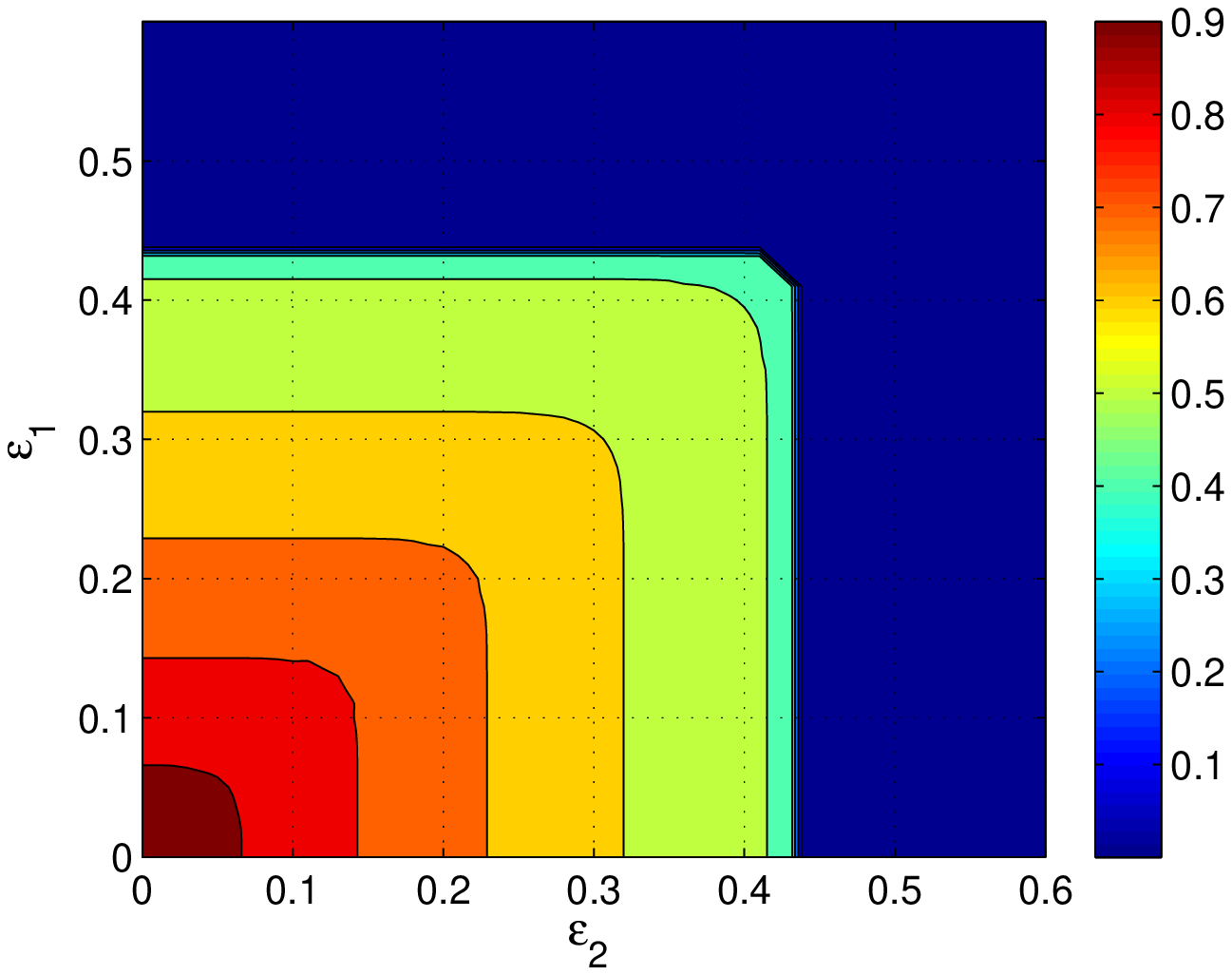}

\label{fig:Rnc}}
\centering{}\caption{Per-link achievable rate region w.r.t. link erasure rate for $\eta_{0}=5\%$,
$r\in[0.5,1]$.\label{fig:PER-LINK CAPACITY}}
\end{figure}

In Fig. \ref{fig:PER-LINK CAPACITY} , we show per-link achievable
rate region of the two schemes w.r.t. per-link erasure rates for different
target $\eta_{0}$ with $r\in[0.5,1]$. For each ($\delta_{1},\delta_{2}$),
we choose a coding rate as large as possible so that $\eta^{e2e}$
and $\eta^{NC}$ satisfy $\eta_{0}$, respectively with end-to-end
coding and NC-case. We term achievable region as the region where
$\eta_{0}$ is satisfied. Otherwise, the region in blue represents
the cases in which there is not any coding rate in the given range
that meets $\eta_{0}$. As denoted in Fig. \ref{fig:R2e2}, it is
interesting to note that each region is shaped by curves. On the other
hand, Fig. \ref{fig:Rnc} denotes $R^{NC}$ w.r.t. link erasure rates
with $\eta_{0}=5\%$ where achievable regions are shaped by squares.
In general, we can conclude that (1) the shapes of $R^{e2e}$ and
$R^{NC}$ are curves and squares, respectively, regardless the constraints
of $r$ and $\eta_{0}$, (2) achievable rate region obtained with
NC-case is almost twice wider than that of end-to-end coding for the
same limitation of coding rate and $\eta_{0}$, and (3) the shapes
of $R^{NC}$ and $R^{e2e}$ are symmetric over the diagonal. $R^{NC}$
and $R^{e2e}$ are independent of the order of the two links. The
proof is not shown here due to lack of space. 

Our numerical results indicate that re-encoding at the intermediate
nodes can significantly enhance network throughput. For the latter,
in the specific application of geo-controlled network reliability,
we thus assume that NC is deployed both at the source and intermediate
nodes along the path whenever an intermediate node is ready to implement
NC function.

\subsection{Geo-controlled network reliability}

We evaluate reliability gain and connectivity gain for the case of
using our design of VGNCF with respect to path length for communication
services beyond the coverage area of satellite in low and high complexity
constraints. Databases with geo-tagged link statistics and geo-location
information are utilized for the optimization functionality towards
the energy-efficient use of network resources.

For illustrative purposes, we assume that a great number of network
devices are uniformly distributed in a deployment area. Complexity
constraints are the same for all nodes, i.e. $\beta_{0}^{S}$=$\beta_{0}^{R_{j}}$=$\beta_{0}^{D}$=$\beta_{0}$.
All links undergo the same erasure rate for different cases of $\delta=0.1$
and $0.15$, $k=50$ information packets, $L=100$ bytes, $q=8$,
$\rho_{0}=80\%$ for two different computational complexity constraints,
$\beta_{0}^{low}=8\times10^{6}$ and $\beta_{0}^{high}=10\times10^{6}$
logic gates. Coding rate is optimized according to Section \ref{sec:Optimization}.
We conduct various numerical results to evaluate how network performance
will be improved with VGNCF according to system limitations and different
network conditions such as erasure rate, path length, etc. 

\subsubsection{Network reliability gain \label{sec:Reliability}}

\begin{figure}[tbh]
\begin{centering}
\includegraphics[scale=0.45]{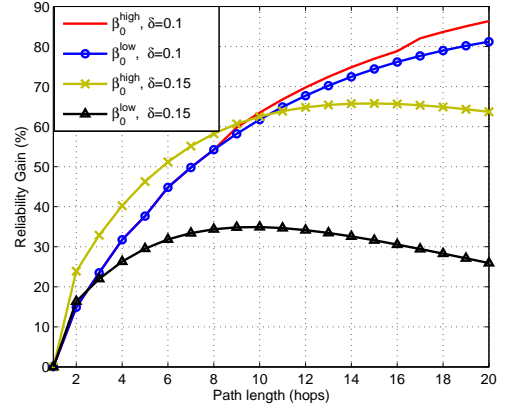}
\caption{Network reliability gain with NC, $\rho_{R}^{NC}-\rho_{R}^{noNC}$,
according to optimized utility for different path length and erasure
rates, with $\rho_{0}=80\%$.\label{fig:RELIABILITY}}
\par\end{centering}
\end{figure}

In Fig. \ref{fig:RELIABILITY}, we show network reliability gain with
NC, i.e. $\rho_{R}^{NC}-\rho_{R}^{noNC}$, according to optimized
utility at the source for different path length and erasure rates
in high and low complexity constraints, respectively. A glance at
the figure reveals that network reliability with NC outperforms that
of the case of transmission without NC. Note that we only denote the
path length up to $20$ hops in which the performance of no-NC case
is extremely low and thus $20$ hops are large enough for our comparison.
In particular, in case of $\beta_{0}^{high}$ and $\delta=0.1$, reliability
gain significantly increases with path length. This is because of
that $\beta_{0}^{high}$ can guarantee the target $\rho_{0}=80\%$
for a path length of more than $20$ hops (even some hundred hops)
in low erasure rate. In contrast, performance without NC degrades
dramatically with the number of hops. For longer path, the proposed
VGNCF just needs to increase the redundant level of combined packets
in order to cope with the erasure process. Otherwise, consider $\beta_{0}^{high}$
and $\delta=0.15$, the target is only satisfied up to approximately
$10$ hops due to the fact that the longer the transmission path,
the lower the connectivity due to physical limits. The limitations
of redundant combined packets then cannot provide the reliability
as required. Even though maximum redundancy is chosen, the utility
function is not possible to reach maximum point.

\subsubsection{Connectivity gain \label{sec:CONNECTIVITY}}

Assuming some reliability design target $\rho_{0}$, in the uncoded
case, many nodes would not achieve the target $\rho_{0}$ while NC
case could. For simplicity, the connectivity gain for a target $\rho_{0}$
is defined as $\gamma\left(\rho_{0}\right)=h^{NC}(\rho_{0})/h^{noNC}(\rho_{0})$
where $h^{NC}(\rho_{0})$ and $h^{noNC}(\rho_{0})$ denote the hop
at which NC and the uncoded case can provide connectivity with the
reliability satisfying $\rho_{0}$, respectively.

\begin{table}[tbh]
\begin{centering}
\subfloat[$\rho_{0}=80\%$]{\begin{centering}
\begin{tabular}{|c|c|c|c|}
\hline 
$\gamma\left(\rho_{0}\right)$ & $\beta_{0}^{very-low}$ & $\beta_{0}^{low}$ & $\beta_{0}^{high}$\tabularnewline
\hline 
\hline 
$\delta=0.1$ & 9 & 32 & 335\tabularnewline
\hline 
$\delta=0.15$ & 2 & 4 & 11\tabularnewline
\hline 
\end{tabular}
\par\end{centering}
}
\par\end{centering}
\begin{centering}
\subfloat[$\rho_{0}=85\%$]{\begin{centering}
\begin{tabular}{|c|c|c|c|}
\hline 
$\gamma\left(\rho_{0}\right)$ & $\beta_{0}^{very-low}$ & $\beta_{0}^{low}$ & $\beta_{0}^{high}$\tabularnewline
\hline 
\hline 
$\delta=0.1$ & 13 & 47 & 495\tabularnewline
\hline 
$\delta=0.15$ & 1 & 3 & 8\tabularnewline
\hline 
\end{tabular}
\par\end{centering}
}
\par\end{centering}
\caption{Connectivity gain (in times) since using VGNCF for different erasure
rates and complexity constraints, with ($a$) $\rho_{0}=80\%$ and
($b$) $\rho_{0}=85\%$. \label{tab:CONNECTIVITY_GAIN}}
\end{table}

In Table \ref{tab:CONNECTIVITY_GAIN}, we depict connectivity gain
when using VGNCF for different $\rho_{0}$ with $\beta_{0}^{very-low}=5\times10^{6}$,
$\beta_{0}^{low}=8\times10^{6}$, and $\beta_{0}^{high}=10\times10^{6}$
logic gates. The larger the computational constraint, the higher the
connectivity gain beyond the cell coverage of the satellite. Particularly,
since $\delta=0.1$, NC can obtain up to $335$ times and $495$ times
gain in connectivity if compared to the uncoded case with high constraint
for $80\%$ and $85\%$ target reliability, respectively. Otherwise,
in low constraint, connectivity gain obtained with NC is $32$ times
and $47$ times for $80\%$ and $85\%$ target reliability, respectively.
The reason is that the performance of the uncoded case is significantly
impacted by link erasure rate and length of transmission path. NC
case, meanwhile, can adapt its coding rate within the constraints
to obtain the target while optimizing source's utility. However, implementing
VGNCF is then strongly affected by computational complexity which
can be equivalently mapped into the price to pay in terms of energy
consumption.

\section{Conclusions \label{sec:CONC}}

In this paper, we have proposed the integration of NC and NFV architectural
design as a toolbox so that NC can be designed as a VNF thus providing
flow engineering functionalities to the network. We have conducted
a complete design to illustrate the use and relevance of our proposed
VGNCF design for reliable communication over satellite, where geographical
information is the key enabler to support VGNCF. Specifically, we
show how our proposed VGNCF should interact with NFV-MANO blocks through
a general procedure for the instantiation, performance monitoring,
and termination of VGNCF. Especially, optimization functionality ensures
an optimized operation of VGNCF. Several numerical results show the
improvement of overall throughput (achievable rate), reliability gain,
and connectivity gain with NC if compared with the uncoded case. Our
proposed framework can naturally be tailored for different designs
and accommodate additional functionalities leading to implementation
and deployment of VGNCF design for next generation networks.

\section*{Acknowledgment}

This research was financially supported by H2020 GEO-VISION - GNSS
driven EO and Verifiable Image and Sensor Integration for mission-critical
Operational Networks (project reference 641451). 

\section*{Appendix 1: Proof of Theorem 1}
\begin{IEEEproof}
(3) follows the min-cut of the network, (4) shows the constraints
in terms of the target RPER at each receiver, and (5) follows the
below argument.

We have $1-\eta^{m}=\Pi_{i=1}^{m-1}(1-\eta_{i})$. Because $1-\eta_{m-1}\le1$,
by multiplying $\Pi_{i=1}^{m-2}(1-\eta_{i})$ for both sides, then
$1-\eta^{m}=\Pi_{i=1}^{m-1}(1-\eta_{i})\le\Pi_{i=1}^{m-2}(1-\eta_{i})$$,$$\forall r,\delta_{i}$.
Hence, 
\begin{equation}
R^{m}=r(1-\eta^{m})\text{\ensuremath{\le}}R^{m-1},\text{\ensuremath{\forall}}r,m\text{\ensuremath{\in}}[2,|V|].
\end{equation}

At an arbitrary point in the capacity, i.e. $\eta_{0}=0,\,\,\eta^{|V|}=0$
iff $\eta_{1}=\eta_{2}=...=\eta_{|V|-1}=0$. Therefore, $R^{sink}=R^{m}=r$.
Then, for $r\in[r_{o},1]\,\,(r_{o}\in[0,1]),$ $r_{o}\le R^{sink}\le1$
and $r_{o}\le R^{m}\le1$. 

At an arbitrary point in the achievable rate region, and the target
$\eta_{0}\neq0,R^{m}<R^{m-1},\forall r,m\in[2,|V|].$ 
\end{IEEEproof}
\bibliographystyle{IEEEtran}
\bibliography{satcom}

\end{document}